\documentstyle[prb,aps,preprint]{revtex}
\draft
\author{Simone Artz and Steffen Trimper}
\address{Fachbereich Physik\\
Martin-Luther-Universit\"at\\D-06099 Halle Germany}
\title{Competing Glauber and Kawasaki Dynamics}
\begin{document}
\tightenlines
\draft
\date{\today}
\maketitle
\begin{abstract}
\noindent Using a quantum formulation of the master equation we study 
a kinetic Ising model with competing stochastic processes: the Glauber 
dynamics with probability $p$ and the Kawasaki dynamics with probability 
$1 - p$. Introducing explicitely the coupling to a heat bath and the mutual 
static interaction of the spins the model can be traced back exactly to 
a Ginzburg Landau functional when the interaction is of long range order. 
The dependence of the correlation length on the 
temperature and on the probability $p$ is calculated. In case that the spins 
are subject to flip processes the correlation length disappears 
for each finite temperature. In the exchange dominated case the system is 
strongly correlated for each temperature.
\\ \\ \\

\end{abstract}
		
\section{Introduction}
\noindent The time evolution of nonequilibrium Ising spin systems is  
of interest in particular when the spins are subject to different 
dynamics \cite{gf,gf1,gf2,ml}. Initiated by Glauber \cite{g}, for a review see 
\cite{k}, a stochastic model had been considered in which each spin can flip 
with a certain transition rate constructed in such a way that detailed balance 
is fulfilled. The system has to be in contact with a heat bath not specified 
in detail. The bath yields the energy for such a 
stochastic flip--process where every spin can flip in a first view 
independently on the orientation of neighbouring spins. 
In general it is assumed that the dynamics of the whole system 
is governed by a stochastic master equation. Recently, a very 
powerful method in analysing master equations was proposed 
\cite{doi,gra,pe,gw,satr,sd,scsa,aldr}. 
It based on a quantum formulation of the underlying master equation 
written in terms of creation and annhiliation operators. Originally, the 
method had been formulated for Bose operators \cite{doi,gra,pe}. 
More appropriate, in particular for problems with an exclusive dynamics, is 
the application of quantum Pauli operators which allows exact 
solutions for a whole class of one dimensional systems, for a review see \cite{sti}.
We have proposed an alternative way to include the coupling to a heat bath 
\cite{sctr} explicitely. The method has been applied to study a 
spin--facilitated model \cite{st} where the dynamics is restricted by specific 
constraints. It can be also used to find the above mentioned transition 
probability \cite{si} which is not unique. It should be emphasized that the 
method is 
determined by the algebraic properties of the operators where an extension to 
a p--state model \cite{sctr2} and to a Q--statistics is straightforward 
\cite{sctr3}.\\
Using such a Fock space formulation of the master equation we analyse 
a model where the system experiments with propability $p$ single spin flips 
(Glauber dynamics \cite{g}) and with propability $1 - p$ 
spin exchange processes (Kawasaki dynamics \cite{k}). The coupling to a heat 
bath with fixed temperature $T$ is included and the static interaction between 
the spins is assumed to be of infinite range \cite{ka}. Whereas the Glauber 
kinetics is always related to a change of the order parameter 
(nonconserved order parameter), the Kawasaki dynamics simulates the flux 
of energy into the system (conserved order parameter). 
In a continuous approach the model consists of a combination of model A 
and model B within the classification proposed by Hohenberg and Halperin 
\cite{hh}.\\
There is a great effort in analysing such nonequilibrium spin 
systems with competing dynamics. Gonzalez-Miranda et al \cite{ggml} had  
studied a kinetic Ising model where Glauber and Kawasaki dynamics drive 
the system simultaneously. Using Monte Carlo simulations in two dimensions 
they found the phase diagram in the plane temperature versus the rate of 
Kawasaki process, here $1 - p$. It reveals a line of continuous transitions 
between the ferro--and the paramagnetic phases. Assuming that the temperature 
depends on $p$ there appears for low temperatures a nonequilibrium 
tricritical point which had been confirmed by Dickman \cite{di} 
employing the dynamic pair approximation. In the present paper, 
the rate $p$ and the temperature $T$ are assumed to be independent variables.\\  
Another interesting feature observed is the occurence of self organization 
phenomena \cite{to}. Increasing the energy flux the system changes 
continuously from the ferromagnetic to a paramagnetic state. A further 
increase of flux drives the system into an antiferromagnetic phase 
which has been confirmed numerically at infinite temperature \cite{gf2} and for 
finite temperatures \cite{gf1}. Based on a Monte Carlo simulation Grandi 
et al \cite{gf} was sucessful in calculating the critical exponents when the 
Kawasaki dynamics dominats the behaviour of the system. An analytical approach 
has recently given by Ma et al \cite{ml}.\\
Here, we are interested in a quantum formulation of the problem using a 
Fock space representation of the master equation. The model can be 
attributed to the standard Ginzburg Landau functional exactly when the static 
interaction is realized by an infinite range model. 
Calculating the correlation function we find the correlation length depending 
on the independent variables temperature $T$ and the rate $p$ which is the probabilty 
that the Glauber dynamics is realized. In the model there is no indication for 
a tricritical point.

\section{Fock space representation}

\noindent The dynamics is govered by the master equation written in the 
symbolic form  
\begin{equation}
\partial_tP(\vec n,t)=L^{\prime}P(\vec n,t) 
\label{ma}
\end{equation}
Here $P$ is the probability that a certain configuration in terms of 
a lattice gas representation, $\vec n = (n_1, n_2 \dots n_N), $ is realized at 
time $t$. The evolution operator $L^{\prime }$ specified below is given by 
competing flip and exchange processes. The variables $n_i$ are related to 
the occupation number operators with eigenvalues $0$ and $1$. Hence, the 
problem is to formulate the dynamics in such a way that this 
constraint is taken into account \cite{gw,satr,sd,scsa,aldr,sti}. The 
situation in mind can be analyzed in a 
seemingly compact form using a Fock space representation of the master 
equation \cite{doi,gra,satr} introduced by eq.(\ref{ma}). 
Following \cite{doi,gra,pe,satr,aldr} the probability distribution 
$P(\vec n,t)$ is related to a state vector $\mid F(t) \rangle$ in a 
Fock-space according to $P(\vec n,t) = \langle \vec n\mid F(t)\rangle$ 
where the basic-vectors $\mid \vec n \rangle$ are composed of second 
quantized operators. 
The master equation (\ref{ma}) can be transformed into an equivalent 
equation in a Fock-space
\begin{equation}
\partial_t \mid F(t)\rangle = L \mid F(t) \rangle
\label{fo1}
\end{equation}
where the operator $L'$ in eq.(\ref{ma}) is mapped onto the operator $L$. 
Up to now the procedure is independent on the used operators. Originally, the 
method had been applied for the Bose case \cite{doi,gra,pe}. Recently, 
an extension to restricted occupation numbers (two discrete orientations) 
was proposed \cite{gw,satr,sd,scsa,aldr} based upon a Pauli-operator 
representation.  
These operators commute at different sites and anti-commute at the same 
lattice site. A further extension to an p--fold occupation number is 
possible \cite{sctr2}.\\
\noindent The relation between the quantum--like formalism and the probability 
approach is given by
\begin{equation}
\mid F(t) \rangle = \sum_{n_i} P(\vec n,t) \mid \vec n \rangle
\label{fo2}
\end{equation}  
It had been shown by Doi \cite{doi} that the average of an arbitrary physical 
quantity $B(\vec n)$ can be calculated using the average of the 
corresponding operator $\hat{B}(t)$ in according to
\begin{equation}
\langle \hat{B}(t) \rangle = \sum_{n_i} P(\vec n,t) B(\vec n) = 
\langle s \mid \hat{B} \mid F(t) \rangle 
\label{fo3}
\end{equation} 
with the state function $\langle s \mid = \sum \langle \vec n \mid$. Using the 
relation $\langle s \mid \hat{L} = 0$ the evolution equation for an operator 
$\hat{B}$ can be written in the form  
\begin{equation}
\partial_t \langle \hat{B} \rangle = \langle s \mid [\hat{B},\hat{L}] \mid F(t) \rangle
\label{kin}
\end{equation}
It should be noted that all dynamical equations govering the 
classical problem are determined by the commutation rules of the underlying 
operators and the structure of the evolution operator $L$. 
In our case the dynamics of the model is given by spin-flip processes 
indicating a change of the local spin orientation and through 
the Kawasaki exchange dynamics.\\  
\noindent The evolution operator for a local single flip--process reads
\cite{sctr} 
\begin{equation}
L_f = \sum_i [\lambda (1 - d_i) d^{\dagger}_i + \gamma (1 - d^{\dagger}_i) d_i]
\label{fli1}
\end{equation}
where $\lambda$ and $\gamma$ are state independent flip rates. 
A generalization to flip processes with restriction is discussed in 
\cite{st}\\
\noindent The operators $d_i$ and $d_i^{\dagger}$ fulfil the 
commutation rule of Pauli--operators. The occupation number operator 
$n_i = d_i^{\dagger} d_i$ is related to the Ising spin variable 
by $S_i = \frac{1}{2} -  n_i$.\\
The evolution operator for spin conserving exchange processes reads
\begin{equation}
L_e = \frac{\tilde{\mu}}{2} \sum_{<i,j>} [ 
(1 - d_i d^{\dagger}_j) d^{\dagger}_i d_j + (1 - d_j d^{\dagger}_i) d^{\dagger}_j d_i ]
\label{ex}
\end{equation}
with the exchange rate $\tilde{\mu}$.\\
\noindent The complete dynamics is given by a superposition
\begin{equation}
L = p L_f + (1 - p) L_e
\label{fex}
\end{equation}
The quantity $p$ represents the probability that the spins 
follow the Glauber dynamics whereas $1 - p$ characterizes the amount of spins 
which are subjected to an exchange process.\\
Up to now the flip or exchange processes may be performed independently on the 
enviroment in which the system is embedded and on the static interaction between 
the spins. The explicit coupling to a heat bath and the inclusion of 
the static interaction can be realized, see \cite{sctr}, by replacing the 
evolution operators $L_f$, eq.(\ref{fli1}), by 
\begin{eqnarray}
L_f = &=& \nu \sum \left[ (1 - d_i) \exp(-\beta H/2) d^{\dagger}_i 
\exp(\beta H/2) \right] \nonumber\\ 
&+& \left[ (1 - d^{\dagger}_i) \exp(-\beta H/2) d_i \exp(\beta H/2) \right] 
\label{fli2}
\end{eqnarray}
Here, $\nu$ is a new hopping rate, $\beta = T^{-1}$ is the 
inverse temperature of the heat bath and $H$ is the Hamiltonian describing 
the static interaction between the spins. In general the inclusion of the 
mutual interaction leads to nonlocal terms already in the flip 
evolution operator eq.(\ref{fli1}).
In case of the exchange dynamics the evolution operator $L_e$, eq.(\ref{ex})
should be rewritten in the form 
\begin{eqnarray}
L_e &=& \frac{\mu}{2} \sum_{<i,j>} \left[ (1 - d_i d^{\dagger}_j) 
\exp(-\beta H/2) d^{\dagger}_i d_j \exp(\beta H/2)  \right]\nonumber\\
&+& \left[ (1 - d_j d^{\dagger}_i) \exp(-\beta H/2) d^{\dagger}_j 
d_i \exp(\beta H/2) \right]
\label{ex1}
\end{eqnarray}
where $\mu$ is the exchange rate.

\section{Infinite range model}
\noindent The thermalization introduced by eqs.(\ref{fli2},\ref{ex1}), 
is complete different to the conventional approach due to Glauber, 
see \cite{k}. The Hamiltonian $H$, specified below by eq.(\ref{ham1}), 
mediates the coupling to the bath at the fixed temperature $T$ and to the mutual 
interaction present also in the static limit. Physically, the replacement 
of the operator $d^{\dagger}_i$ by $\exp(-\beta H/2) d^{\dagger}_i \exp(\beta H/2)$ 
in (\ref{fli2}) means that a flip--process is realized with a weighting rate 
$\exp(\beta H/2)$. After performing the flip, manifested by $d^{\dagger}_i$,  
the final state is related to the weighting rate $\exp(-\beta H/2)$ consistent 
with the fact that only single flip--processes are taken into account. 
Hence, the procedure simulates in an analytical manner Monte Carlo steps. Due 
to the thermalization a spin flip is not independent on the 
orientation of the other spins. Instead the process is self organized by the 
static coupling of the spins themselves.\\ 
As the simplest case we consider an infinite range model introduced by 
Kac \cite{ka}. By reason of the long range interaction the model exhibits an 
exact static solution according to the mean field approximation. 
The Hamiltonian is defined by 
\begin{equation}
H = -\frac{J}{2N} \sum_{1\le i < j \le N} S_i S_j
\label{ham1}
\end{equation}
Using this Hamiltonian in the evolution operator eq.(\ref{fex}) there 
appears a conflicting situation between the local spin flip process and the 
infinite range interaction manifested by eq.(\ref{ham1}). Whereas the 
long range static interaction favours a reordering of many spins 
the dynamics is refered to single spin flip processes. The situation is 
comparable with those known from spin glasses where the local 
disorder leads to restrictions within the dynamical processes. 

\section{Dynamical equation}
\noindent Using the algebraic properties of Pauli--operators the thermalized 
evolution operators $L_f$ (\ref{fli2}) is rewritten as:
\begin{eqnarray}
L_f &=& \nu \sum_i [\lambda (1 - d_i) d^{\dagger}_i e^{-E} + 
(1 - d^{\dagger}_i) d_i e^{E}] \nonumber\\
\mbox{with} \quad E&=& \frac{J}{2TN} \sum_j S_j
\label{fli5}
\end{eqnarray}
Different to the short range model (nearest neighbour coupling) 
the energy $E$ depends on the values of all other spins. 
It should be noted, that $E$ is obtained if terms with 
$i = j$ are not taken into consideration in eq.(\ref{ham1}). When the 
diagonal terms are regarded the 
dynamical approach does not reproduce the correct stationary state. Within a 
static consideration the mentioned exclusion is irrelevant.\\
As the result the flip process is a thermal activated one realized with a 
certain rate depending on the energy $E$ including the temperature. 
Within the infinite range model it is easy to derive that the exchange 
process is not thermalized. The corresponding terms are canceled out mutually.\\ 
Taking into account the analytical form of $L_f$ and $L_e$ and using 
eq.({\ref{kin}), the evolution equation for the order parameter can be obtained:
\begin{equation}
\partial_t \langle S_r \rangle = \nu p 
(\langle \sinh E \rangle - 2 \langle S_r \cosh E \rangle) + \mu (1-p) 
\sum_{l(r)} \langle S_l - S_r \rangle
\label{gl}
\end{equation}
where $l(r)$ means summation over all nearest neighbours of the lattice site 
$r$. Note that the first part, proportional to $p$, is originated from the 
flip process where the second one, proportional to $1 - p$, can be 
attributed to the exchange dynamics. The situation is different to the 
conventional 
analysis where the flip processes contribute to spatial correlations, too.\\    
The evolution equation for the higher order terms appearing in eq.(\ref{gl}) 
could be calculated in a 
straightforward manner, see \cite{si}. However, in the large $N$--limit 
those terms has not to be evaluated provided the system is an ergodic one. It 
results 
\begin{equation}
\lim_{N \to\infty} E = \langle E \rangle = \frac{J}{2T} \langle S_r \rangle
\label{en}
\end{equation}
Consequently, eq.(\ref{gl}) becomes a closed equation for the order 
parameter itself. In the vicinity of the phase transition the present approach 
leads to a Ginzburg Landau functional, see for instance \cite{hh}. 
In a continuous approximation for $\varphi(\vec x,t) = \langle S_r(t) \rangle$ 
it reads
\begin{eqnarray}
\partial_t \varphi &=& \nu p \left( \sinh(\epsilon \varphi) - 
2 \varphi \cosh(\epsilon \varphi) \right) + \mu (1-p) \nabla^2 \varphi \nonumber\\ 
\mbox{with}\quad \epsilon &=& \frac{J}{2T} \equiv \frac{2 T_c}{T} 
\label{gi}
\end{eqnarray}
The correlation function can be calculated within an expansion in terms of the 
order parameter $\langle S_r(t) \rangle$. In lowest order we find 
\begin{equation}
C_0(\omega, \vec k) \simeq \frac{1}{ -i \omega + k^2 + \xi^{-2}}
\label{cor}
\end{equation}
The correlation length $\xi$ depends on the concentration $p$ and 
$T - T_c$ in the following manner 
\begin{equation}
\xi = \sqrt{\frac{1-p}{p}} \xi_0 \quad \mbox{with} \quad \xi^2_0 \simeq 
\frac{l^2\mu}{2\nu(1-T_c/T)} 
\label{cor1}
\end{equation}
The result can be easily interpreted. If the probability $p =1$ all 
spins are subject to a stochastic flip--process. But, due to the long range 
interaction, static fluctuations are complete suppressed. 
The long range force tends to align all spins parallel preventing 
spatial fluctuations. Consequently, the system is not able to establish 
spatial correlations, hence the correlation length is zero. In the opposite 
case $p \to 0$, when the whole spins are subject to an exchange dynamics, 
all neighbouring pairs of spins tend to fix their orientation 
(spin conservation). Then, the fluctuations are extremly strong and for $p = 0$ 
the system remains in a strongly correlated state for finite temperatures. As 
the result the correlation length goes to infinity 
for each temperature. In the intermediated 
region $ 0 < p < 1$ the correlation length is a product of a $p$--dependent 
part and a temperature--dependent part originated from the Ginzburg Landau 
functional. As already emphasized, the situation is complete different to 
the conventional approach where the energy $E$ is a local energy leading 
to additonal spatial fluctuating terms originated from the spin--flip terms. 
One can also estimate that the inclusion of higher order fluctuations does 
not change the situation. A renormalization group procedure based on a 
Ginzburg Landau expansion starts with the correlation function $C_0$ 
obtained in (\ref{cor}). Such an approach yields only a 
modified critical exponent for the correlation length $\xi_0$ in eq.(\ref{cor1}).
There is also no indication for a tricritical point.

\section{Conclusions}
\noindent In this paper we have traced back a model with 
competing Glauber and Kawasaki dynamics to a Ginzburg Landau functional where 
the correlation length depends explicitely on the amount of spins which 
undergo single flip processes and on the temperature. This result could be 
achieved by a coupling to a heat bath and by the consideration of the mutual 
static interaction between the spins. When this interaction is of long range 
order we obtain a closed set of equations for the order parameter. As the 
result of several conflicting situations, long range static force versus 
a single spin flip process or nearest neighbour exchange coupling, or 
Glauber versus Kawasaki dynamics the correlation length of the system 
depends on the probability $p$ which is a measure of the amount of spins 
which are subject to a Glauber process. It should be emphasized that the 
model is neither model A nor model B in the conventional classification 
\cite{hh}. Different to this classification the Ginzburg Landau 
functional consists of two parts, a nonconserving and spatial independent 
part and a diffusive (conserving) one originated by the exchange dynamics.

\end{document}